\documentclass[12pt]{iopart}

\usepackage{graphicx}
\usepackage{amssymb}
\usepackage{amsfonts}
\usepackage{algorithm}
\usepackage{algorithmic}
\usepackage{dcolumn}
\usepackage{bm}
\usepackage{bbold}
\usepackage{float}
\usepackage{color}
\newcommand{\ignore}[1]{}
\begin{document}


\title{Irreversibility and Entanglement Spectrum Statistics in Quantum
  Circuits}

\author{Daniel Shaffer$^1$, Claudio Chamon$^1$, Alioscia Hamma$^2$ and
  Eduardo R. Mucciolo$^3$} 

\address{$^1$Department of Physics, Boston University, Boston,
  Massachusetts 02215, USA}

\address{$^2$Center for Quantum Information, Institute for
  Interdisciplinary Information Sciences, Tsinghua University, Beijing
  100084, P.R. China}

\address{$^3$Department of Physics, University of Central Florida,
  Orlando, Florida 32816, USA}


\begin{abstract}
  We show that in a quantum system evolving unitarily under a
  stochastic quantum circuit the notions of irreversibility,
  universality of computation, and entanglement are closely
  related. As the state evolves from an initial product state, it gets
  asymptotically maximally entangled. We define irreversibility as the
  failure of searching for a disentangling circuit using a
  Metropolis-like algorithm. We show that irreversibility corresponds
  to Wigner-Dyson statistics in the level spacing of the entanglement
  eigenvalues, and that this is obtained from a quantum circuit made
  from a set of universal gates for quantum computation. If, on the
  other hand, the system is evolved with a non-universal set of gates,
  the statistics of the entanglement level spacing deviates from
  Wigner-Dyson and the disentangling algorithm succeeds. These results
  open a new way to characterize irreversibility in quantum systems.
\end{abstract}



\section{Introduction}

The onset of irreversibility in physics is one of the great questions
at the heart of statistical mechanics. The second principle of
thermodynamics in essence states that spontaneous processes happen in
one direction, namely increasing entropy or decreasing free
energy. Boltzmann's H-theorem was aimed in proving this possibility
starting from microscopic time-reversal mechanics, but, as he
tragically had to realize, there is no way of doing that. In classical
physics irreversibility either has to assume some seed of randomness,
or it must resort to coarse graining and topological mixing
\cite{terhaar1955}. Of course, in classical physics, coarse graining
and the counting of micro states are arbitrary operations, and it was
soon recognized that firmer grounds for statistical mechanics must be
found in the quantum domain.

At a first glance, though, the quantum case looks even harder. In a
closed system evolution is unitary, and therefore the entropy of a
quantum state cannot increase. Moreover, unitary evolution is always
reversible, so irreversibility is strictly speaking impossible. In
classical mechanics irreversibility is due to chaos, that is, high
sensitivity to initial conditions. But in quantum mechanics, unitarity
implies that slightly different initial conditions do not evolve into
highly different states.

Recently, \cite{popescu2006}, irreversible phenomena like
thermalization have been interpreted in terms of entanglement. Because
of entanglement, every subsystem of a pure quantum system will
typically increase its entropy to give thermal expectation values to
most of its observables. However, it is not clear exactly in what way
entanglement is involved in the irreversible character of
thermalization. In fact, also non thermalizing (integrable) systems
increase their entropy to saturation. After one has stated that
irreversibility is due to entanglement, we are left with the feeling
that one still needs to specify what kind of entanglement. What notion
of irreversibility one should use, is part of the question, as
well. Moreover, one would like to tackle all these questions in a more
general setting than evolution governed by a Hamiltonian. We mentioned
above that irreversibility in quantum mechanics cannot happen at the
level of wave function. If two initial conditions are slightly
different, their overlap in time will remain bounded. This is a direct
consequence of unitarity. Instead, one can consider the Loschmidt
echo, that is, the overlap between the state obtained by evolving the
initial state with some Hamiltonian $H$ and then reversing the
evolution with a slightly perturbed Hamiltonian $H+\epsilon V$:
$\mathcal L =|\langle\psi (0)|e^{i(H+\epsilon
  V)t}e^{-iHt}|\psi(0)\rangle|^2$. A fast decay of $\mathcal L$ is
what is called irreversibility~\cite{Peres84}. It is remarkable that
quantum irreversibility and non-integrability are not as closely
connected as one may think \cite{benenti2009}, which is another sign
of the non satisfactory conceptual standing of the whole matter.

In this paper, we take a completely new point of view. We take
seriously the idea that the defining feature of quantum mechanics is
entanglement. Irreversibility and thermalization must be a consequence
of entanglement. As we shall see, it is not the amount of entanglement
per se that is important. This is too much of a simple and generic
notion. A specific pattern of entanglement, though, is what is
important for irreversibility. This approach offers several
advantages. First of all, it is novel. Questions that have been asked
for decades and remained somehow unresolved can now be looked from a
new point of view. In particular, entanglement is not neutral (like
energy) with respect to the degrees of freedom that are
involved. Entanglement is defined within a chosen tensor product
structure, that is, it depends on what the local observables are
\cite{zanardi2004}. Understanding thermalization must mean finding
which observables thermalize, {\it i.e}, understanding what the
relevant tensor product structure is. We posit that if the evolution
is irreversible the entanglement will be complex in the degrees of
freedom in which the interactions are simple. In this sense
irreversibility is a complexity problem. Every Hamiltonian is
diagonalizable, but simple interactions may result in complex
entanglement. The second advantage is that this approach is apt to
study any quantum evolution. One does not need a Hamiltonian, and not
even unitary evolution, in principle. Every evolution that takes a
quantum density matrix into another quantum density matrix ({\it i.e},
a completely positive map), that is, any quantum process, can be
studied with this method.

The method, first introduced in Ref. \cite{PRL}, is easily
explained. We evolve the state with a quantum evolution, starting from
a completely disentangled state. After a while, the state will be
fully entangled. Then we try to go back to the initial state, that is,
we try to undo the entanglement. To this end, we use a Metropolis-like
algorithm. We select a random local unitary (that is, involving up to
two degrees of freedom), and check whether this decreases some measure
of entanglement. If yes, we keep it, and we draw another random
unitary operator. If not, we reject it. After a certain number of
local unitaries, we may have two outcomes: (i) the state is fully
disentangled, or, (ii) no matter how long we keep drawing local
unitaries, the state remains entangled. In the case (i) we have
reversed the evolution trying out a small number of unitaries. Since
disentangling the state is easy, we say that entanglement is not
complex. On the other hand, the case (ii) is irreversible. We see that
irreversibility means the incapability of disentangling the state,
unless one tries out all the possible evolutions. Irreversible
entanglement is thus complex entanglement.

At this point, we ask ourselves what makes some entanglement complex
or not. If the entanglement is complex, one should be able to look at
the final state and find a pattern that tells us that there is no way
to disentangle it in a simple way. To this end, we look at the level
spacing statistics of the reduced density matrix spectrum, namely, the
eigenvalues of the reduced density matrix of a subsystem. We show that
the irreversible, complex entanglement corresponds to a Wigner-Dyson
statistics of the eigenvalue spectral fluctuations, showing a robust
level repulsion. On the other hand, simple, removable entanglement
possesses spectral fluctuations that follow, for instance, a Poisson
statistics, where no level repulsion is present. Therefore, a generic
quantum (unitary) evolution is irreversible because of a specific
pattern of entanglement. As we explained, irreversibility is thus
understood in terms of the complexity of the entanglement, not in
terms of how much a state is entangled.

In this paper, we push forward the ideas first proposed in
Ref. \cite{PRL} where this phenomenon was noticed for a specific
subset of quantum circuits, namely, those generated by the subgroup of
the unitary group realized by the permutation subgroup of unitary
transformations. However, the question was raised whether the findings
in Ref. \cite{PRL} are just a property of the particular
transformations chosen, or if, instead, the notion of irreversibility
based on entanglement makes sense for general quantum unitary
evolutions. To this end, we consider a quantum system made of $N$
qubits, and subject to a random unitary quantum circuit. The local
unitaries are one- and two-qubit gates. The set containing the gates
{\bf H} (Hadamard), {\bf T} (phase gate), and {\bf CNOT} (controlled
NOT) is a universal set of gates, meaning that every unitary can be
decomposed in a circuit made of these. If any unitary can be produced,
every state in the Hilbert space can be reached. Of course, also {\bf
  CNOT} plus a random one-qubit unitary is universal. We find that
evolution obtained with the full unitary group by means of universal
quantum computation is irreversible and the statistics of the level
spacings of reduced density matrix spectrum is Wigner-Dyson, {\it
  i.e}, is complex. On the other hand, if the entanglement level
spacing statistics is Poisson, we find that the disentangling
algorithm succeeds, which means that the pattern of entanglement is
not complex. Strikingly, this happens when the set of gates used is
not universal for quantum computing. It is remarkable that
reversibility arises with severe breaking of ergodicity. If the set of
gates is not universal, not every state can be reached, and ergodicity
is broken.

\section{Entanglement in stochastic unitary quantum circuits}

We consider discrete quantum mechanical systems made of qubits, that
is, two level systems with Hilbert space $\mathcal H_i\simeq \mathbb
C^2\simeq\mbox{span}\{|0\rangle,|1\rangle\}$, the latter being a
orthonormal basis. The total Hilbert space of $n$ qubits is thus
$\mathcal H =\otimes_{i=0}^{n-1} \;\mathcal H_i$, that is spanned by
$2^n$ vectors. A basis can be conveniently written in terms of tensor
product states $|k\rangle = \bigotimes^{n-1}_{i=0} |x_i \rangle \equiv
|x_{n-1} ... x_0\rangle$ where $x_i = 0$ or $1$. $k = x_{n-1} ... x_0$
is just a number expressed in binary digits; this is therefore
referred to as the numerical basis. A general state of the system is a
linear combination with complex numbers of these basis states,
$|\Phi\rangle = \sum_{k=0}^{2^n-1}\phi_k|k\rangle$. Normalization of
probability imposes that $\sum_{k=0}^{n-1} |\phi_k|^2=1$. On a classical
computer, such a state can be represented by a complex vector with
$2^n$ elements: $(\phi_0, \dots , \phi_{2^n-1})$.

The unitary evolution of the state is performed by applying quantum
gates. Any unitary operation can be performed by a sequence of gates
acting on one or two qubits at a time. Furthermore, only a finite
number of such gates is needed to form the universal set of
operations.

Here we consider the following sets of gates.

$\bullet$ {\bf NOT} gate: flips the state of the qubit,
$|0\rangle\to|1\rangle$ and $|1\rangle\to|0\rangle$.

$\bullet$ {\bf H} (Hadamard) gate: takes $|0\rangle\to \frac{1}{\sqrt
  2}\left(|0\rangle+|1\rangle\right)$ and $|1\rangle\to \frac{1}{\sqrt
  2}\left(|0\rangle-|1\rangle\right)$.

$\bullet$ Phase gates {\bf P}$_\delta$: gives a state dependent phase,
$|0\rangle\to|0\rangle$ and $|1\rangle\to e^{i\delta}|1\rangle$. The
phase gate with $\delta=\pi/4$ is called {\bf T} and that with
$\delta=\pi/2$ is called {\bf S}.

$\bullet$ {\bf CNOT} gate or controlled-NOT gate: this is a two-qubit
gate that performs a NOT on a controlled (second) bit if the control
(first) bit is 1 and does nothing if the control bit is 0:
$|00\rangle\to|00\rangle$, $|01\rangle\to|01\rangle$,
$|10\rangle\to|11\rangle$, $|11\rangle\to|10\rangle$.

The set $\mathcal I$, comprising, {\bf H}, {\bf P}$_\delta$, and {\bf
  CNOT} is sufficient to reach arbitrarily close to any state in the
Hilbert space, provided $\delta$ is such that $\theta =
2\cos^{-1}\left(\cos^2(\delta/2)\right)$ is an irrational multiple of
$\pi$, for example when $\delta = \frac{\pi}{4}$ (as in the {\bf T}
gate) \cite{nielsenchuang}. The condition on $\theta$ can be
understood from the requirement that the whole Bloch sphere must be
accessible, and since unitary operations correspond to rotations of
the Bloch sphere, this is only possible if a rotation by an irrational
angle is constructable from the set of operations. In contrast, the
set of gates {\bf H}, {\bf S}, and {\bf CNOT} does not lead to
universal quantum computing; this set spans a subset of operations,
the Clifford group \cite{Gottesman}. Although one can express
arbitrary unitaries on the space of $n$ qubits in terms of a handful
of types of gates, the number of operations needed to implement any
given unitary gate may be very large, in principle countably infinite.

The inverse of any unitary operation written using a sequence of {\bf
  H}, {\bf P}$_\delta$, and {\bf CNOT} gates is simply the sequence
written backwards, with {\bf P}$^{\;}_{-\delta}=${\bf
  P}$^{-1}_{\delta}$ in place of {\bf P}$^{\;}_{\delta}$ in the
reverse sequence. That such a reverse sequence exists, with exactly
the same length as the original sequence, says nothing about the
difficulty of finding the reverse computation if the sequence is
concealed.

The statement of universality of a set of gates $\mathcal I$ means
that any unitary operator $U\in \mathcal U(\mathcal H)$ can be well
approximated as the product of $U_{a}\in\mathcal I$, that is, $U\simeq
\prod_{a=1}^K U_{a}$. The unitary $U$ is called {\em quantum
  circuit}. In the following, we consider stochastic quantum
circuits. A stochastic quantum circuit of $K$ gates is obtained by
randomly selecting $K$ gates $U_{a}\in \mathcal I$, as well as the
either one or two qubits on which such gates act.

We are interested in studying several entanglement properties of
states that are obtained by evolving with a stochastic quantum circuit
$U$. The initial state is a pure state $|\Psi\rangle$ with the
completely factorized form
\begin{equation}
\label{eq:Psi_0}
|\Psi_0\rangle = |\psi_0\rangle \otimes |\psi_1\rangle \otimes \cdots
|\psi_{n-1} \rangle,
\end{equation}
where
\begin{equation}
\label{eq:psi_j}
|\psi_j\rangle = \cos(\theta_j) |0\rangle_j + \sin(\theta_j) |1\rangle_j
\end{equation}
Being $|\Psi_0\rangle$ completely factorized, every restriction of the
state to any subsystem $A$ is still a pure state. On the other hand,
if $|\Psi_0\rangle$ were entangled, we would obtain a classical
probability distribution for a mixed state. Entanglement is relative
to the partitioning of the system in two parts $A,B$, that is, by
considering the Hilbert spaces $\mathcal H_A, \mathcal H_B$ of
$n_A,n_B$ given qubits such that $\mathcal H=\mathcal H_A\otimes
\mathcal H_B$, where $n=n_A+n_B$. Given a (pure) state
$|\psi\rangle\in\mathcal H$, the reduced density matrix $\rho_A$
associated to the restricted system $A$ is obtained by tracing out the
qubits in $B$:
\begin{equation}
\rho_A = \mbox{tr}_B |\psi\rangle\langle\psi|
\end{equation} 
The eigenvalues of $\rho_A$, that is, $\mbox{spec}\{ \rho_A\}=
\{\lambda_1,...,\lambda_{2^{n_{^{\!A}}}}\}$, form the reduced density
matrix spectrum. They are indeed just the spectrum of the Hermitian
operator $\rho_A$. The operator $\rho_A$, being a density matrix, has
unit trace, so $\sum_{i=1}^{2^{n_A}}\lambda_i=1$, with
$\lambda_i\ge0$, thus representing a classical probability
distribution. In the following, we are interested in studying the
spectrum of the reduced states $\rho_A^U$ after the quantum circuit
$U$ has been applied on the initial fiducial state $|\Psi_0\rangle$:
\begin{equation}
\rho_A^U = \mbox{tr}_B \rho^U =\mbox{tr}_B\left(
U|\Psi_0\rangle\langle\Psi_0|U^\dagger\right)
\end{equation} 

The entropy measures of the probability distribution $\{\lambda_i\}$
associated to $\rho_A^U $ are given by the R\'enyi entropies
$S_\alpha$, namely
\begin{equation}
S_q (\rho_A^U)= \frac{1}{1-q}\log_2\mbox{tr}(\rho_A^U)^q
\end{equation}
In the limit $q=1$, this is the von Neumann entropy of entanglement,
while $S_0$ is the number of nonzero eigenvalues of $\rho_A^U$, {\it
  i.e.}, its rank. Of course, for a factorized state all R\'enyi
entropies $S_q=0$ for all $q$. Moreover, we recall the useful
inequality $S_q\ge S_{q'}$ for $q<q'$. Notice that the R\'enyi
entropies depend explicitly on the choice of the subsystem $A$ and on
the circuit $U$. So, for every circuit $U$ and every partition $A$, we
obtain a set of quantities $S_\alpha (\rho_A^U)$. As we mentioned
above, the stochastic quantum circuit $U$ is made of $K$ gates, so we
may denote it by $U{(K)}$. Here, $K$ plays the role of time. At every
click of the clock, we apply a new gate and make the circuit
longer. Given a number $K$ and a partition $A$, one can make many
realizations of the quantum circuit $U{(K)}$. Then, it makes sense to
consider the averages of the R\'enyi entropies over the many circuit
realizations, namely $S(q,K,A)\equiv\overline{S_q
  (\rho_A^U(K))}^U$. Again, this quantity depends explicitly on
$K$. As $K$ increases, one expects $\overline{S_q (\rho_A^U(K))}^U$ to
start at zero and monotonically increase to the maximum possible value
of the entropy, as it was shown in Ref. \cite{hamma2012}. After we
reach a state with maximum entanglement for a given system size $n$,
we can try to disentangle it without resorting to the inverse of
circuit $U(K)$, namely, $U(K)^\dagger$. Of course, by exhaustive
search, one can try all possible quantum circuits for a given set of
gates until we find the right one. This search is very expensive and,
already for some dozens of qubits, it is prohibitive. If there is
nothing better than brute search, the entanglement is too {\em
  complex} to be undone, and we say that the evolution is
irreversible. On the other hand, if there is an efficient way of
finding a circuit $U(K')$ such that $U(K')U(K)|\Psi_0\rangle$ is
completely factorized, then the entanglement is not complex, and the
evolution is reversible. The algorithm used to find the disentangling
algorithm resembles the Metropolis algorithm to reach the ground state
of a quantum system, and we thus dub it {\em entanglement
  cooling}. The algorithm will be explained in the next section.

At this point, it is desirable to characterize the entanglement in the
state $U(K)|\Psi_0\rangle$ {\em without} having to try to disentangle
it. Is there anything in the entanglement itself that shows us,
beforehand, whether the entanglement cooling will work or not? As
shown in Ref. \cite{PRL}, the answer is hidden in the statistics of
the level spacing for the entanglement spectrum. To be precise, we
order the members of $\mbox{spec}(\rho^U_A)$ in decreasing order, and
consider the distribution of the ratio of consecutive spacings,
namely, $P(r) = \frac{1}{R} \sum_{i=1}^R \langle
\delta(r-r_i)\rangle$, where $r_i = \epsilon_{i+1}/\epsilon_{i}$ and
$\{\epsilon_{i} =\lambda_i-\lambda_{i+1}\}_{i=1,\ldots,R}$. As we
shall show, irreversibility is associated to a Wigner-Dyson statistics
for this distribution. Moreover, if $U$ is obtained by $\mathcal I$,
this is always the case. On the other hand, if the quantum circuits
are not drawn from a universal set, the entanglement cooling algorithm
always works, entanglement is not complex, and the distribution $P(r)$
deviates from Wigner-Dyson prediction and is akin to a Poisson
one. This is the main result of the paper.

\section{Simulation Setup}

In this section, we describe the detail of the simulations and
results. Our system comprises $16$ qubits. The simulations are made of
three parts: entanglement heating, entanglement cooling, and level
spacing statistics.

\subsection{Entanglement heating}

\begin{figure}[ht]
\begin{center}
\includegraphics[width=1 \textwidth]{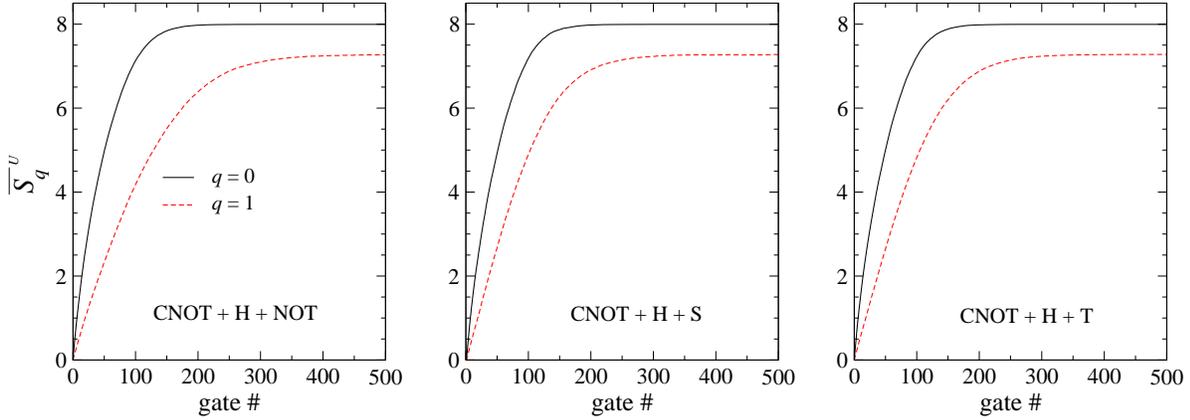}
\caption{Average entanglement entropies $\overline{S_0}^U$ (solid
  line) and $\overline{S_1}^U$ (dashed line) as a function of gate
  number for three circuit cases.  The cases {\bf CNOT+H+NOT} and {\bf
    CNOT+H+S} are reversible, while the case {\bf CNOT+H+T} is
  not. Notice that the entanglement entropies themselves do not flag
  which circuit is easily reversed or not. The complexity of the
  entanglement is instead contained in the reduced density matrix
  spectrum level statistics presented below.}
\label{fig:ave_entr}
\end{center}
\end{figure}

We refer to the initial evolution by the stochastic random circuit $U$
as {\em entanglement heating}, since, as we shall see, we start from a
completely factorized state until reaching maximum entanglement, as
one would expect from a random quantum circuit \cite{hamma2012}. The
circuit $U$ is obtained by drawing random gates from different
sets. The set $\mathcal I$, as mentioned above, is universal and
comprises \textbf{H}, \textbf{T}, and \textbf{CNOT}. The sets used are
therefore
\begin{eqnarray}
\quad \quad \mathcal I & = & \{CNOT + H+T\} 
\nonumber \\
{\rm or}\quad
\mathcal I & = & \{CNOT +NOT+ H\} 
\nonumber \\
{\rm or}\quad
\mathcal I & = & \{CNOT + H+S\} \nonumber
\end{eqnarray}
The initial state $|\Psi_0\rangle$ is always a (random) factorized as
explained in the previous section. The random state is chosen by
randomizing the $\theta_j$ in Eq.(\ref{eq:psi_j}) with uniform
probability over the interval $[0:\pi]$. In the simulation, we first
randomly select a qubit, then with uniform probability choose what
gate to apply (and, in the case when a two-qubit gate is picked, the
second qubit is then randomly chosen as well). The heating circuit $U$
is made of $512$ gates. From the graph theoretic point of view, we
have a random quantum circuit on a complete graph. All heating
simulations are conducted with $5000$ realizations. For every
realization of $U$, we compute $S_0(\rho_A^U), S_1(\rho_A^U)$ where
$A$ is a subset with $8$ qubits. Then we average over all the
realizations, generating $\overline{S_0(\rho_A^U)}^U$ and
$\overline{S_1(\rho_A^U)}^U$. In Fig. \ref{fig:ave_entr} we show the
average R\'enyi entropies as a function of the number of gates applied
for the three circuit cases. For this plot,≈ß≈π the entropy was
obtained after bipartitioning the bit string into two equal parts,
with $A$ corresponding to bits 0 to 7 and $B$ corresponding to bits 8
to 15. Notice that while $\overline{S_0}^U$ always reach its maximum
possible value after a few hundred gates, the saturation value of
$\overline{S_1}^U$ varies from circuit to circuit and is not a good
indication of whether the circuit is reversible or not by the cooling
algorithm. The maximum saturation value is expected to happen in the
thermodynamic limit. We clearly see that the different sets of gates
are not distinguished at all by the way entanglement increases and
saturates. In particular, entanglement is not able to tell apart
universal from non-universal set of gates.

\subsection{Entanglement cooling and level spacing statistics}

In our simulation, we start with a random product state, so the
entanglement entropy increases from zero to a maximum value after a
certain number ($\sim n^2$) of gates. We therefore refer to this part
of the process as entanglement heating. After a state reaches the
maximum entanglement observed for that system size, our task is thus
to disentangle it, {\it i.e.}, reduce the entanglement entropy to zero
without using knowledge of the random circuit. As alluded to above,
the procedure is a Metropolis-like algorithm: we apply a gate on the
entangled state at random, chosen from the same set of gates from
which the random circuit was constructed; if the entanglement entropy
decreases, we keep the gate; if it increases, we reject the gate. The
process goes on until the entanglement entropy is zero, or until we
give up. In order to reach the completely factorized state, we
actually consider the average of particular bipartitions $(A,B)$ of
the system with $n_A=n-n_B$ running from $1$ to $n-1$. These
bipartitions are constructed by laying down the qubits on a line, and
taking the $n-1$ possible cuts; system $A$ is always chosen to be the
qubits on the left, and $B$ those on the right. Also, more generally
than simply rejecting the gates that increase the entanglement, we
actually use a simulated annealing type variant, where we accept the
increase in entropy with a probability $\exp[-\beta(S_{\rm new} -
  S_{\rm old})]$, with $\beta$ chosen as $\beta=5$ in units of
entropy. We refer to this part of the simulation as entanglement
cooling. The algorithm is summarized below.

\begin{algorithm}
\caption{Entanglement Cooling Algorithm}
\algsetup{indent=2em}
\begin{algorithmic}[1]
\STATE $S_{\rm old} \leftarrow \mbox{entanglement\ entropy\ of\ } \psi$
\WHILE{$S_{\rm old} > 0$}
\STATE Apply gate at random: $\psi \leftarrow \psi_{\rm new}$
\STATE $S_{\rm new} \leftarrow $ entanglement entropy of $\psi$
\IF{$S_{new} > S_{\rm old}$}
\STATE $r \leftarrow x \in [0,1]$
\IF{$r > \exp[\-\beta(S_{\rm new} - S_{\rm old})]$}
\STATE Undo gate: $\psi \leftarrow \psi_{\rm old}$
\ENDIF
\ENDIF
\STATE $S_{\rm old} \leftarrow S_{\rm new}$
\ENDWHILE
\end{algorithmic}
\end{algorithm}

The cooling algorithm is computationally heavy on a classical machine
because it involves a singular value decomposition. Furthermore, it is
necessary at each step to compute not just a single entanglement
entropy, but entanglement entropies over all possible ordered
bipartitions. This is due to the fact that even if the state is a
product of two substates, the two substates may themselves be
entangled. On a quantum machine, the algorithm could be accelerated by
replacing the entanglement entropy with the second R\'enyi entropy,
which is an observable quantity and so could be used during
computation \cite{PRL}. Such performance boosts of classical
algorithms is the main thrust behind the development of a quantum
computer.

Not having a quantum machine, however, we may want to know whether the
algorithm will fail before we attempt it. In that case, we would
probably not want to run it on a quantum machine either. The point of
the proposed conjecture is that it does not require any computation to
be performed on the state; the state itself is all that is needed to
obtain the reduced density matrix spectrum. After heating up the
state, and before we attempt to cool it, we therefore look at its
entanglement spectrum. Because the heating process is fast, we can
obtain spectra over many realizations of the system. If we observe
level repulsion, {\it i.e.}, a Wigner-Dyson level spacing behavior, we
predict that the system will not be reversed by the cooling algorithm.

In practice, evaluating the level spacing distribution directly is
difficult because of the enormous variations in the level density. In
Ref. \cite{PRL}, an elaborate unfolding procedure was used to make the
reduced density matrix spectrum uniform. Here, to avoid this
procedure, we evaluate the distribution of ratios of adjacent level
spacings. An analysis of spectral fluctuations based on such
distribution was first used (for the case of energy, but not
entanglement levels) in studies of finite-temperature many-body
localization by Oganesyan and Huse \cite{oganesyan07}. Later, very
accurate surmises for this distribution were derived by Atas and
coworkers \cite{atas13} for the Gaussian ensembles. They are given by
the expression
\begin{equation}
P_{\rm WD}(r) = \frac{1}{Z} \frac{\left(r+r^2\right)^\beta}
{\left(1+r+r^2\right)^{1+3\beta/2}},
\label{eq:Pr_WD}
\end{equation}
where $Z=\frac{8}{27}$ for the Gaussian Orthogonal Ensemble (GOE) with
$\beta=1$, and $Z=\frac{4}{81}\frac{\pi}{\sqrt{3}}$ for the Gaussian
Unitary Ensemble (GUE) with $\beta=2$. The corresponding distribution
for a spectrum with Poisson statistics is
\begin{equation}
P_{\rm Poisson}(r) = \frac{1}{(1+r)^2}.
\label{eq:Pr_P}
\end{equation}
Notice that $P_{\rm WD}(r\rightarrow 0) \sim r^\beta \rightarrow 0$,
which indicates level repulsion, while $P_{\rm Poisson}(r\rightarrow
0) \rightarrow 1$. The tails of these distributions are also markedly
different, with $P_{\rm WD}(r\rightarrow \infty) \sim 1/r^{2+\beta}$
while $P_{\rm Poisson}(r\rightarrow \infty) \sim 1/r^2$.

For the ``cooling'' simulations, out of the 5000 samples, we picked
randomly 100 samples and attempted to use the disentangling algorithm
to return their maximally entangled states back to a product state
form. The restricted number is due to the fact that the cooling
algorithm is very time consuming. The results of the simulation are
presented below. However, for the statistical analysis, we used all
5000 samples generated for each circuit set.

For each circuit case, we show a typical evolution of the entanglement
entropy as gates are applied, both during the ``heating'' and
``cooling'' processes. Other plots show the resulting eigenvalue level
spacing ratio distribution collected at the end of the heating
process. To highlight the behavior of the distributions at the tails,
we show the data in both linear and logarithm scales.

\subsubsection{\bf CNOT + H + NOT}

\begin{figure}[ht]
\centering
\includegraphics[width=0.5 \textwidth]{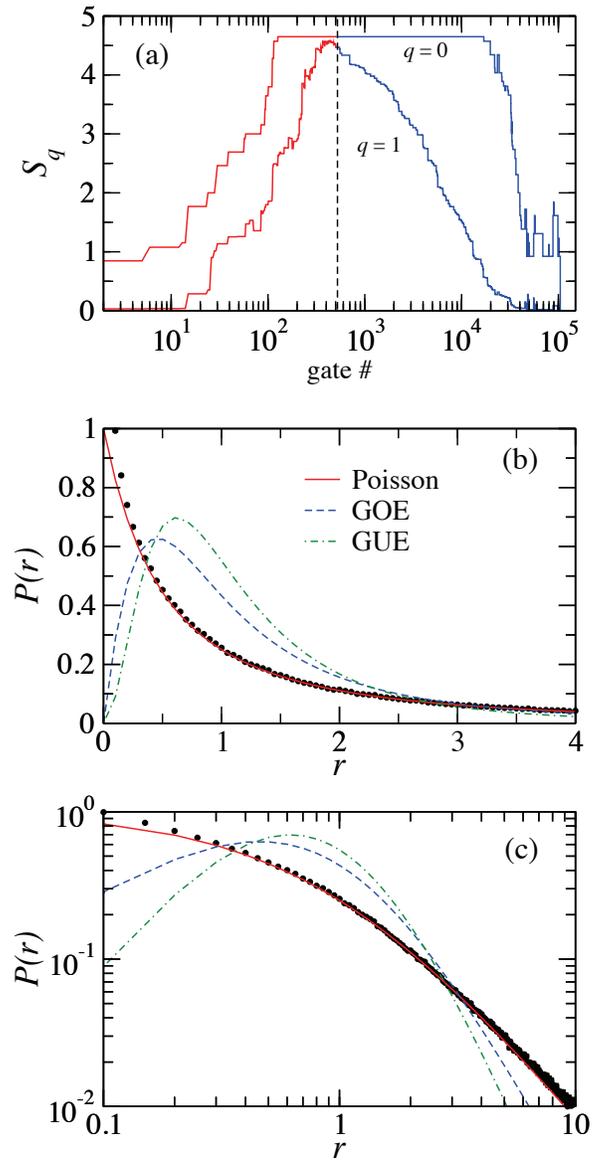}
\caption{CNOT + H + NOT circuit. (a) Sample evolution of the
  entanglement entropies $S_0$ and $S_1$ (averaged over $n-1$
  bipartitions) as functions of gate number. The dashed line separates
  the heating and cooling processes (notice that the gate number is
  given in logarithmic scale). (b) and (c): distribution of eigenvalue
  spacing ratios in linear and logarithmic scales, respectively. The
  solid circles represent the numerical data while the solid, dashed,
  and the dotted lines are the Poisson, GOE, and GUE predictions,
  respectively.}
\label{fig:cnot+h+not}
\end{figure}

This ensemble of gates is not universal. For all 100 samples tested,
the disentangling algorithm was able to brought the state back to a
direct product form. See, for example, the run shown in
Fig. \ref{fig:cnot+h+not}(a). The spacing ratio distribution follows
closely a Poisson curve, as shown in Figs. \ref{fig:cnot+h+not}(b,c).

\subsubsection{\bf CNOT + H + S}

\begin{figure}[ht]
\centering
\includegraphics[width=0.5 \textwidth]{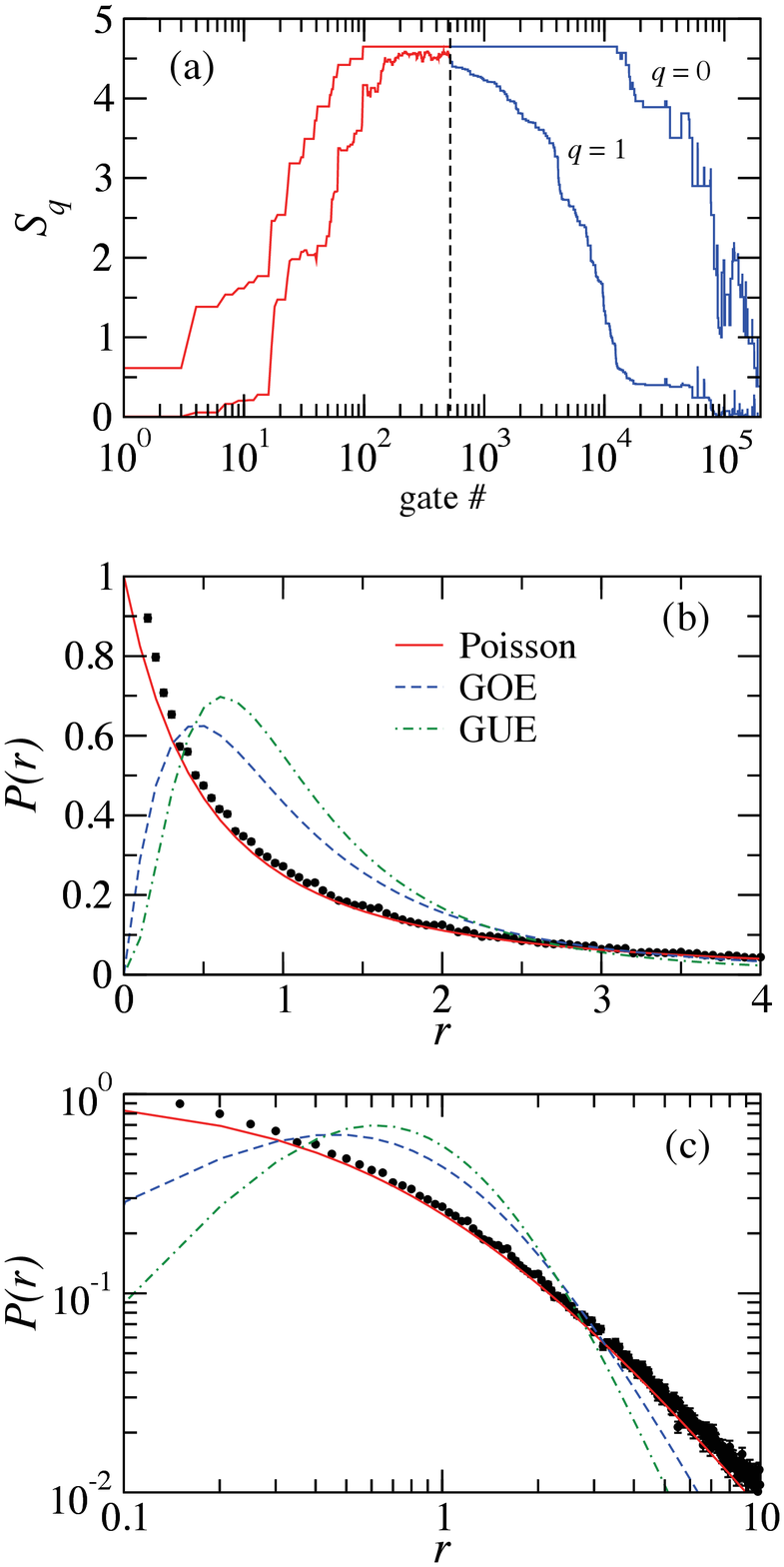}
\caption{CNOT + H + S circuit. (a) Sample evolution of the
  entanglement entropies as functions of gate number. (b) and (c):
  Distribution of eigenvalue spacing ratios (see caption of
  Fig. \ref{fig:cnot+h+not}).}
\label{fig:cnot+h+s}
\end{figure}

As in the previous case, this set is not
universal~\cite{Gottesman}. The cooling algorithm is able to
systematically reduce both $S_0$ and $S_1$ entropies with no tendency
for saturation, although sometimes we were not able to reach the zero
value within the maximum number of steps used (200000). In
Fig. \ref{fig:cnot+h+s} we show a case where nearly a complete
disentanglement was obtained. The distribution of spacing ratios
follows closely a Poisson curve, as indicated in
Figs. \ref{fig:cnot+h+s}(b,c). We can therefore establish that Poisson
statistics correspond to reversibility of the stochastic quantum
evolution.

\subsubsection{\bf CNOT + H + T}

\begin{figure}[ht]
\centering
\includegraphics[width=0.5 \textwidth]{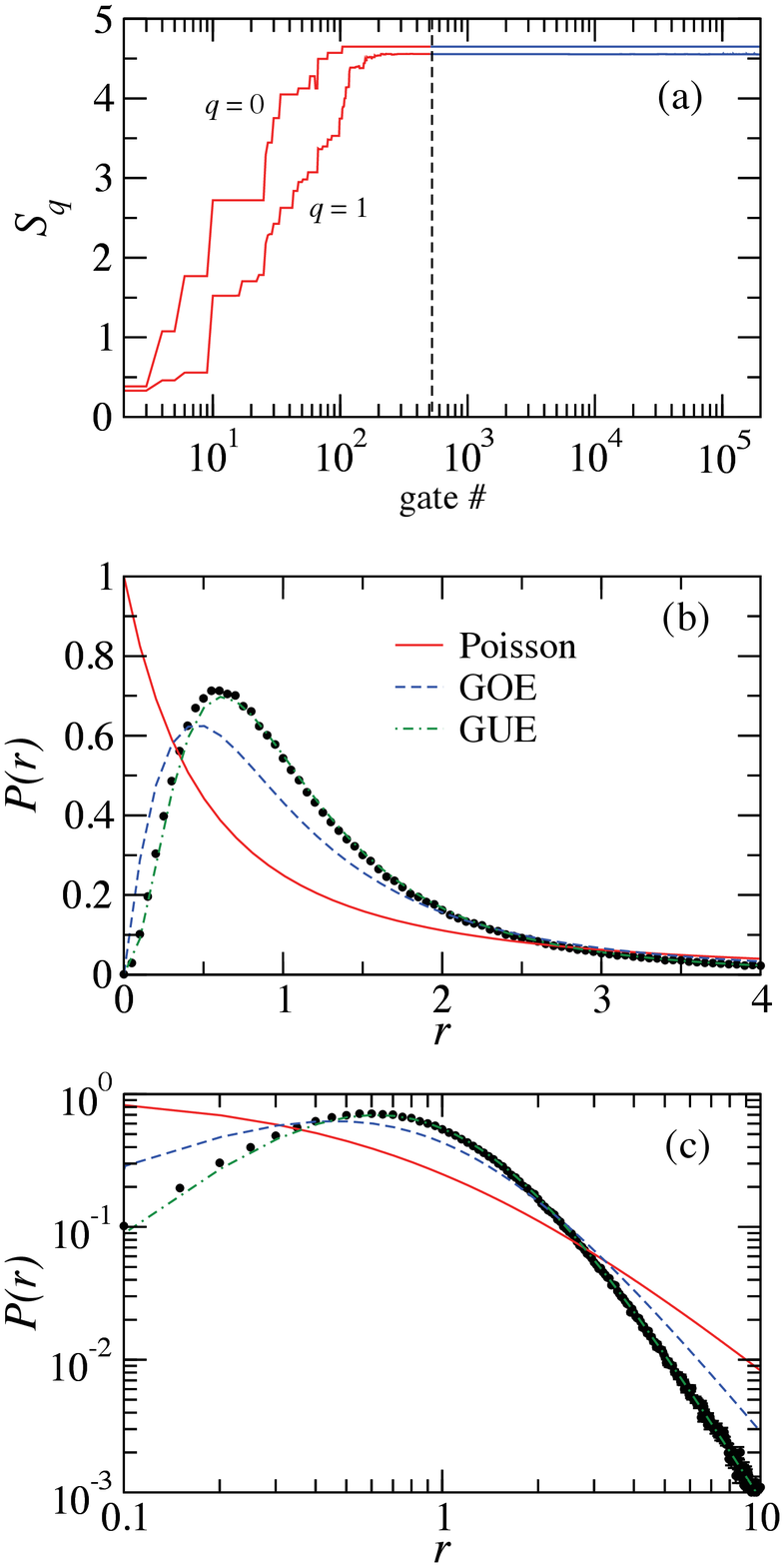}
\caption{CNOT + H + T circuit. (a) Sample evolution of the
  entanglement entropies as functions of gate number. (b) and (c):
  Distribution of eigenvalue spacing ratios (see caption of
  Fig. \ref{fig:cnot+h+not}).}
\label{fig:cnot+h+t}
\end{figure}

This is the universal set of gates $\mathcal I$. We find that for all
100 samples tested, the disentangling algorithm was not only unable to
brought the state back to a direct product, but not even to
disentangle the state partially, as can be seen in the instance shown
in Fig. \ref{fig:cnot+h+t}(a). Notice that in this case neither $S_0$
nor $S_1$ are reduced by the cooling algorithm. Figures
\ref{fig:cnot+h+t}(b,c) show that the distribution of spacing ratios
in this case follows the GUE prediction.

\section{Conclusions and outlook}

In this paper, we have shown that irreversibility and chaotic behavior
in quantum mechanics corresponds to a complex pattern of entanglement
that cannot be undone by entanglement cooling. This pattern is
revealed by Wigner-Dyson statistics in the entanglement level spacing
statistics. We showed that universal quantum computation always
produces complex entanglement. Universal quantum computation means
that every state in the Hilbert space can be reached, so complexity of
entanglement and irreversibility correspond to ergodicity. On the
other hand, non-universal quantum computation results in a breaking of
ergodicity and this is revealed in Poisson level spacing statistics,
in turn ensuring that entanglement is not complex and easily undone.

The results are summarized in the table below. Universal computation
is always irreversible:

\vspace{5mm}
\begin{center}
\begin{tabular}{| l | m{2cm} | c |}
	\hline
	Operator Set & Level Spacing Statistics & Reversible? \\ \hline
	$\{CNOT, H, NOT\}$ & \multicolumn{1}{c|}{Poisson} & Yes \\ \hline
		$\{CNOT, H, S\}$ & \multicolumn{1}{c|}{Poisson} & Yes \\ \hline
		$\{CNOT, H, T\}$ & \multicolumn{1}{c|}{GUE} & No \\ \hline
	
\end{tabular}
\end{center}
\vspace{5mm}

It is important to emphasize that it is not the amount of entanglement
in itself that counts. Typically (that is, it is almost always the
case) quantum evolution brings about the maximum possible
entanglement. But just as some knots are easy to undo and some are
not, this happens to entanglement as well. The complexity of
entanglement is revealed by the level spacing statistics. We speculate
that signatures of this behavior may be found in higher moments of the
entanglement entropy, since after all, knowledge of all the R\'enyi
entropy should in principle allow to reconstruct the whole reduced
density matrix spectrum.

Another point that is worth emphasizing is that this way of looking at
quantum evolutions is very general. We do not need a Hamiltonian. This
means that generic quantum evolutions (even non-unitary, like in open
quantum systems) can be explored in the same way. All we need is the
density matrix. \ignore{It is remarkable that the density matrix does
  know whether the evolution was ``integrable'' or not. In fact, we
  are proposing a novel way of looking at non-integrability. This is
  not a property of the energy spectrum, or of the ``solvability'' of
  the Hamiltonian, but of the associated evolution that may or not
  produce a complex entangled state.}

The results in this work open a way to characterize, through
entanglement, quantum systems that do not thermalize, as breaking of
thermalization must come from breaking of ergodicity. Therefore, long
lived quantum memories or quantum states that feature many-body
localization, should feature deviations from Wigner-Dyson statistics
in their entanglement level spacing statistics. We believe that this
study can shed new light on questions regarding quantum many-body
physics away from equilibrium, and help clarify the notions of
thermalization and irreversibility.

\ack This work was supported in part by the NSF grants CCF-1116590 and
CCF-1117241, and by the National Basic Research
  Program of China Grant 2011CBA00300, 2011CBA00301, the National
  Natural Science Foundation of China Grant 61033001, 61361136003.


\end{document}